\documentstyle[12pt]{article}

\def\MSB{{\rm \overline{MS\kern-0.14em}\kern0.14em}}
\input{psfig}

\catcode`\@=11
\long\def\@makefntext#1{
\protect\noindent \hbox to 3.2pt {\hskip-.9pt
$^{{\ninerm\@thefnmark}}$\hfil}#1\hfill}		

\def\@makefnmark{\hbox to 0pt{$^{\@thefnmark}$\hss}}  

\def\ps@myheadings{\let\@mkboth\@gobbletwo
\def\@oddhead{\hbox{}
\rightmark\hfil\ninerm\thepage}
\def\@oddfoot{}\def\@evenhead{\ninerm\thepage\hfil
\leftmark\hbox{}}\def\@evenfoot{}
\def\sectionmark##1{}\def\subsectionmark##1{}}

\renewcommand{\thefootnote}{\fnsymbol{footnote}}

\newcounter{sectionc}\newcounter{subsectionc}\newcounter{subsubsectionc}
\renewcommand{\section}[1] {\vspace*{0.6cm}\addtocounter{sectionc}{1}
\setcounter{subsectionc}{0}\setcounter{subsubsectionc}{0}\noindent
	{\normalsize\bf\thesectionc. #1}\par\vspace*{0.4cm}}
\renewcommand{\subsection}[1] {\vspace*{0.6cm}\addtocounter{subsectionc}{1}
	\setcounter{subsubsectionc}{0}\noindent
	{\normalsize\it\thesectionc.\thesubsectionc. #1}\par\vspace*{0.4cm}}
\renewcommand{\subsubsection}[1]
{\vspace*{0.6cm}\addtocounter{subsubsectionc}{1}
	\noindent {\normalsize\rm\thesectionc.\thesubsectionc.\thesubsubsectionc.
	#1}\par\vspace*{0.4cm}}

\newcounter{appendixc}
\newcounter{subappendixc}[appendixc]
\newcounter{subsubappendixc}[subappendixc]

\renewcommand{\appendix}[1] {\vspace*{0.6cm}
        \refstepcounter{appendixc}
        \setcounter{figure}{0}
        \setcounter{table}{0}
        \setcounter{equation}{0}
        \renewcommand{\thefigure}{\Alph{appendixc}.\arabic{figure}}
        \renewcommand{\thetable}{\Alph{appendixc}.\arabic{table}}
        \renewcommand{\theappendixc}{\Alph{appendixc}}
        \renewcommand{\theequation}{\Alph{appendixc}.\arabic{equation}}
        \noindent{\bf Appendix \theappendixc #1}\par\vspace*{0.4cm}}

\def\abstracts#1{{

\centering{\begin{minipage}{12.2truecm}\footnotesize\baselineskip=12pt\noindent
	\centerline{\footnotesize ABSTRACT}\vspace*{0.3cm}
	\parindent=0pt #1
	\end{minipage}}\par}}

\renewenvironment{thebibliography}[1]
	{\begin{list}{\arabic{enumi}.}
	{\usecounter{enumi}\setlength{\parsep}{0pt}
\setlength{\leftmargin 1.25cm}{\rightmargin 0pt}
	 \setlength{\itemsep}{0pt} \settowidth
	{\labelwidth}{#1.}\sloppy}}{\end{list}}

\newcounter{itemlistc}
\newcounter{romanlistc}
\newcounter{alphlistc}
\newcounter{arabiclistc}

\newcommand{\fcaption}[1]{
        \refstepcounter{figure}
        \setbox\@tempboxa = \hbox{\footnotesize Fig.~\thefigure. #1}
        \ifdim \wd\@tempboxa > 6in
           {\begin{center}
        \parbox{6in}{\footnotesize\baselineskip=12pt Fig.~\thefigure. #1}
            \end{center}}
        \else
             {\begin{center}
             {\footnotesize Fig.~\thefigure. #1}
              \end{center}}
        \fi}

\newcommand{\tcaption}[1]{
        \refstepcounter{table}
        \setbox\@tempboxa = \hbox{\footnotesize Table~\thetable. #1}
        \ifdim \wd\@tempboxa > 6in
           {\begin{center}
        \parbox{6in}{\footnotesize\baselineskip=12pt Table~\thetable. #1}
            \end{center}}
        \else
             {\begin{center}
             {\footnotesize Table~\thetable. #1}
              \end{center}}
        \fi}

\def\@citex[#1]#2{\if@filesw\immediate\write\@auxout
	{\string\citation{#2}}\fi
\def\@citea{}\@cite{\@for\@citeb:=#2\do
	{\@citea\def\@citea{,}\@ifundefined
	{b@\@citeb}{{\bf ?}\@warning
	{Citation `\@citeb' on page \thepage \space undefined}}
	{\csname b@\@citeb\endcsname}}}{#1}}

\newif\if@cghi
\def\cite{\@cghitrue\@ifnextchar [{\@tempswatrue
	\@citex}{\@tempswafalse\@citex[]}}
\def\citelow{\@cghifalse\@ifnextchar [{\@tempswatrue
	\@citex}{\@tempswafalse\@citex[]}}
\def\@cite#1#2{{$\null^{#1}$\if@tempswa\typeout
	{IJCGA warning: optional citation argument
	ignored: `#2'} \fi}}

 1
 1
 1

\font\ninerm=cmr9

\begin{document}

\centerline{\normalsize\bf QCD FROM THE LATTICE}
\baselineskip=16pt

\centerline{\footnotesize CHRIS MICHAEL}
\baselineskip=13pt
\centerline{\footnotesize\it D. A. M. T. P., University of Liverpool}
\baselineskip=12pt
\centerline{\footnotesize\it Liverpool L69 3BX, U. K.}
\centerline{\footnotesize E-mail: C.Michael@liv.ac.uk}

\vspace*{0.9cm}
\abstracts{
 The lattice gauge theory technique for non-perturbative calculations in
QCD is reviewed.  The extraction of the continuum limit  of lattice
results is discussed with particular examples appropriate  to hadron
spectroscopy (the light hadrons and the glueballs). The determination of
the strong coupling  constant   is presented. The lattice
approach to  the evaluation of hadronic matrix elements appropriate to
weak decays  is described.
}

\normalsize\baselineskip=15pt
\setcounter{footnote}{0}
\renewcommand{\thefootnote}{\alph{footnote}}
\section{Introduction}

 The lattice approach to QCD is, in principle, an exact and  fully
non-perturbative method to extract results. It is an  essential
component in the comparison of predictions of   the Standard Model
with experiment. There are limitations at present, of course, and I aim
to describe the scope and potential of  lattice QCD.

 The lattice approach to QCD, pioneered by Wilson, involves  assigning
the colour fields to the links of a lattice~\cite{text}.  I shall
concentrate on  the use of a four-dimensional lattice with Euclidean
time. Then,  using periodic space-time boundaries, the functional
integral of the quantum field  theory becomes a multi-dimensional
integral over the finite number of  degrees of freedom. Monte Carlo
methods allow this integral to be  performed to the required precision.
This approach thus allows a  completely finite formulation of QCD.
Furthermore, the approach is  fully non-perturbative.  Thus, in
principle, it appears that QCD  is ``solved''.

 Even though  we have an exact non-perturbative method, it is  important
to check the method carefully. The most delicate point is to  confirm
that the limit of zero lattice spacing is under control. This  approach
to the continuum limit will be the first topic I discuss in detail. As
an illustration, I will discuss glueball and light hadron spectra.
The glueball case is discussed because it is an intriguing manifestation
of a possibility in QCD which transcends the quark model. The light
hadron spectrum is discussed to show that lattice methods are quantitative
in a case where the experimental data are available.

 One quantity which forms a link between perturbative and
non-perturbative  QCD is the running coupling $\alpha$. An accurate
knowledge  of $\alpha$ is essential to explore possible contributions
from supersymmetric particles to the unification of electro-weak and
strong couplings.  In principle $\alpha$ can be determined  very
accurately by lattice methods. I review the current state of lattice
determinations.

As well as checking experimentally known quantities,  lattice QCD is
relied on to calibrate  the hadronic component of measurements that
search for physics beyond the Standard Model.  Thus weak decays of
hadrons containing strange, charmed or  bottom  quarks are a source of
data to determine  the CKM mixing matrix and thence  to look for
possible interactions which do not agree with the Standard Model.
Because the Standard Model is defined in terms of the interaction of
quarks  whereas the experiments are performed with hadrons, an accurate
knowledge  of hadronic matrix elements is needed. Lattice QCD can
determine such matrix  elements directly and can also be used to refine
theoretical models for them.

 In principle, lattice QCD provides {\it ab initio} computations of QCD.
The only  inputs are the quark masses and one dimensionful observable to
set the  scale. In practice, computational limitations arise. One of
the main limitations is in dealing with light quarks. Consider a  quark
of mass $m_q$. Fermions are treated  by integrating  out the quark
contributions analytically which yields a large sparse fermionic matrix
$M$.  The computation then involves estimation of the inverse of this
matrix $M$. For heavy quarks, this inverse can be evaluated reliably
and quickly by iteration. For very light quarks (ie with physical masses
of u and d quarks), this iteration is unreliable. The computational
strategy is then to evaluate quantities for a range of quark masses $m_q$
(corresponding to the strange quark mass and heavier) and to extrapolate
to the required light quark masses.

 Because of this limitation of dealing with light quarks, a very popular
 compromise is to use the ``quenched approximation''. In this
approximation,  the sea-quark mass is taken as very heavy so that quark
loops in  the vacuum are neglected. The valence quark masses are treated
as described  above by extrapolation to the required light quark mass
value. As  I shall describe, the quenched approximation seems to
reproduce fairly well the  light quark spectrum and is thus a useful
model for QCD.

 One may go one step beyond the quenched  approximation by explicitly
evaluating the contribution of a  quark loop. This enables
studies~\cite{loops}  of topics such as the $\eta$, $\eta'$ system,
{\it s}-quarks in the nucleon and glueball  decay.

 To go to full QCD, the sea-quark mass must be taken finite and an
extrapolation must be made to physical light quark masses.  Because of
the computational overhead of this study, only limited exploration  has
been made: for example the lattice spacing is usually kept rather
coarse. Some relevant results will  be discussed. This is an area where
future progress is needed.

\subsection{The Continuum Limit}

 Lattice computations use a finite lattice spacing (conventionally
called $a$).  We require $a$ to  be smaller than the overall size of a
hadron and small enough to reproduce sufficiently  high energy
fluctuations in the fields.   Current lattice simulations use a lattice
spacing $a$ ranging from 0.07 fm to 0.20 fm. This is a very reasonable
range given our knowledge of the size of hadrons (over 0.5 fm) and
the energy  scale of the dynamics since distances below 0.2 fm
correspond to energies above 1 GeV. However, it is essential to
quantify the discretisation error from using a non-zero lattice spacing.
I begin by discussing computations in the quenched approximation.

 The discretisation error arising from a given lattice formulation  can
be calculated theoretically. For the simplest discretisation of the
gauge action, that proposed by Wilson, this is  known to be an order
$a^2$ contribution to a dimensionless ratio of observables. Thus if a
mass ratio is measured for several values of $a$,  a reliable
extrapolation can be made to $a=0$. This is illustrated in fig~1.
Clearly an accurate extrapolation to the continuum limit can be made
from the combined data shown for the $0^{++}$ glueball mass. The
earliest results obtained were at rather  coarse lattice spacings
because of computational limitations (ie those on the right on the
figure) and taken alone they  led to  a value of the $0^{++}$  mass
which was too small.

In order to extract a continuum mass, it is appropriate to study  a
dimensionless ratio to the quantity which  is most accurately
determined in  lattice simulation. The potential $V(R)$ between
static quarks is accurately measured on a lattice. Usually the energy
scale from the potential is specified by $\sqrt{K}$ where $K$ (or
$\sigma$) is the  string tension given by $\lim_{R \to \infty} dV/dR$.
Because this  definition involves an extrapolation in $R$, it is more
precise~\cite{sommer} to set  the scale by using the potential at finite
$R$ and choosing an energy  scale $r_0^{-1}$ given implicitly by $R^2
\left. dV/dR\right|_{r_0} = 1.65$. In practice these scales are closely
related since $\sqrt{K} r_0 \approx 1.18$.

In order to give an estimate of the continuum mass in GeV, one needs to
specify a value for $r_0$ (or $\sqrt{K}$) in GeV.  If all predictions of
the quenched approximation were to agree with experiment,  this would
be straightforward. Although many  mass ratios determined in the quenched
approximation do agree with experiment, since some quantities seem to
depart from  the experimental ratios by 10 -- 20\%, it is prudent to
retain an overall  systematic error of 10\% on the energy scale. From
the potential model  of $b\bar{b}$ mesons, one obtains experimental
values for these  scales and to be specific I take  $\sqrt{K}=0.44(4)$
GeV which corresponds to $r_0^{-1}=0.37(4)$ GeV.

\subsection{Glueballs}

\begin{figure}[t]
\vspace{13.6cm}  
\includegraphics{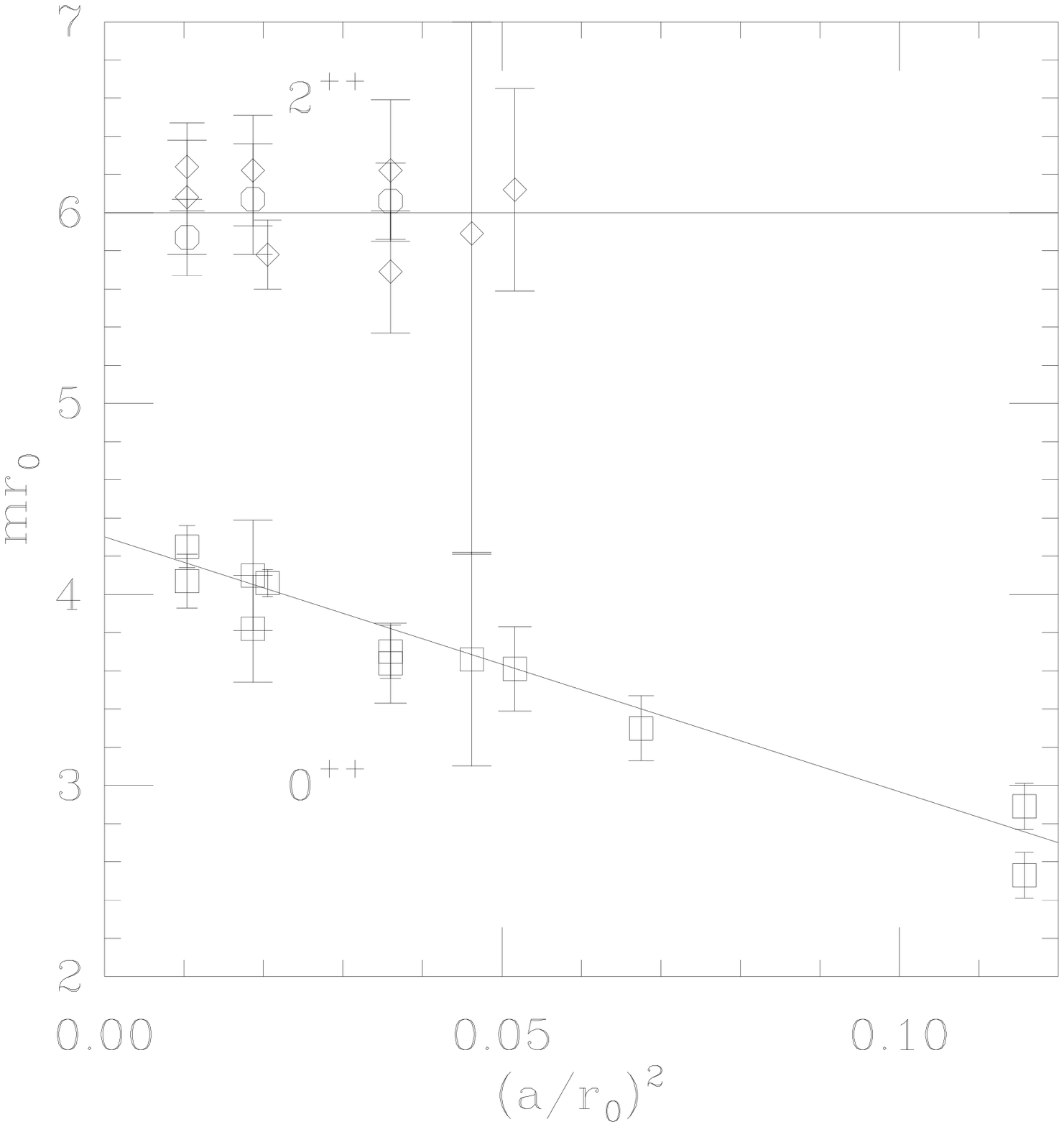}
\fcaption{ The  mass of the  $J^{PC}=0^{++}$ and $2^{++}$
glueball states from refs{\protect\cite{DForc,MT,glue,gf11gl}}
in units of $r_0$.  The straight lines  show  fits describing the
approach to the continuum limit as $a \to 0$.
    }
\end{figure}

 The $0^{++}$ and $2^{++}$ glueball masses in terms of $r_0$ are shown
in  fig~1. Using the theoretical input that these ratios have
corrections  of order $a^2$,  the limits as $a \to 0$ are rather well
determined. Since  the errors on individual lattice determinations are
usually dominated by  systematic errors, some care is needed in fitting
the combined results of different groups.
A reasonable approach  leads to continuum ratios of $m(0^{++})r_0=4.3(2)$
and  $m(2^{++})r_0=6.0(4)$.  Then, using the scale determined as
discussed above,  the lattice glueball  data yield $0^{++}$ and $2^{++}$
masses of 1.6(2) and 2.2(4) GeV  respectively. This systematic error in
the energy  scale can only be removed by going beyond the quenched
approximation.  Then, however, glueball--meson mixing will give problems
in glueball identification  which will be similar to those in
experiments.

 The lattice predictions~\cite{MT,glue} for glueballs of other $J^{PC}$
values are that they lie  higher in mass, except for the $0^{-+}$ state
which appears to  be at a similar mass to the $2^{++}$. There are no
low-lying ``odd-ball'' states  with $J^{PC}$ values not allowed by  the
quark model.

\subsection{Light hadron spectrum}

\begin{figure}[t]

\vspace{11.6cm}  
 \includegraphics{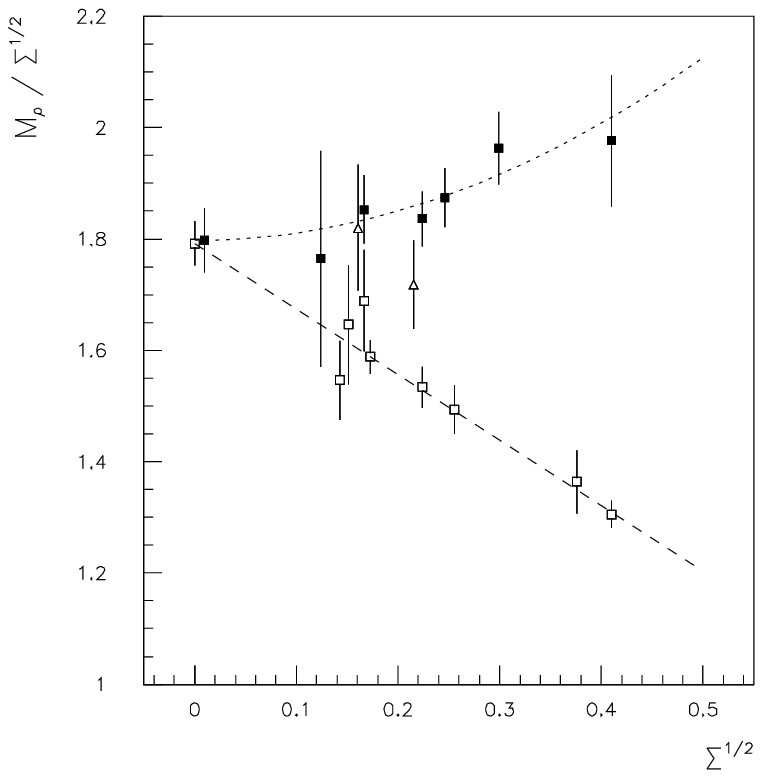} 
\fcaption{
 The mass of the $\rho$ meson as a dimensionless ratio to  the lattice
string tension $\Sigma=Ka^2$. The continuum limit corresponds  to $a \to
0$ at the left hand side; the right hand side corresponds  to $a \approx
0.2 $ fm.  The quenched lattice data with different  fermionic
discretisations are compared~{\protect\cite{soml94}}. The filled squares
are with  staggered fermions, the open squares are from Wilson fermions
and  the triangles are from Clover fermions. The lines illustrate the
discretisation  errors which behave as $a^2$, $a$ and $\alpha a$
respectively.
 }
\end{figure}

There are several  alternative formalisms for describing fermions on a
lattice and I  will show that they agree in the continuum limit.  I
first discuss the results in the quenched approximation  with various
valence quark masses $m_q$. As $m_q$ is decreased,   the pseudoscalar
meson mass is found to extrapolate to zero while the vector  meson mass
tends to a finite limit: to be identified with the mass of  the
$\rho$-meson. Thus the quenched approximation does show  spontaneous
chiral  symmetry breaking.

The dimensionless ratio of the $\rho$ mass to $\sqrt{K}$ is shown by a
compilation~\cite{soml94}  of lattice data in  fig~2 versus the lattice
spacing $a$. The dashed and dotted curves  show that a common continuum
limit exists of $m_{\rho}/\sqrt{K} = 1.80(5)$ from the different
fermionic formulations presented.  Moreover, using the experimental
values  of  the string tension (namely $\sqrt{K}=0.44$ GeV as discussed
above) and  $\rho$-meson mass leads  to  $m_{\rho}/\sqrt{K} = 1.75 $ which
is consistent within the errors with the continuum limit of the lattice
results. So in this case the quenched ratio agrees with the experimental
ratio.

The mass ratio of the proton to $\rho$ is widely known to have a  tendency
to come out too large in quenched lattice calculations. Sceptics  may be
unimpressed by getting a value closer to the naive quark model ratio  of
1.5 than to the experimental value of 1.22. The present situation is
that lattice data {\it are} consistent with the  experimental value when
an extrapolation to the continuum limit is made: see fig~3. A further
extrapolation to infinite spatial volume is also  needed and there is
some  evidence~\cite{gf11} that this can further improve the agreement.
Although the data are consistent with the experimental value, further
reduction in the  errors of the lattice results is needed to answer
definitely whether the quenched approximation  reproduces this mass
ratio or not.

\begin{figure}[t]
\vspace{12.5cm} 
\includegraphics{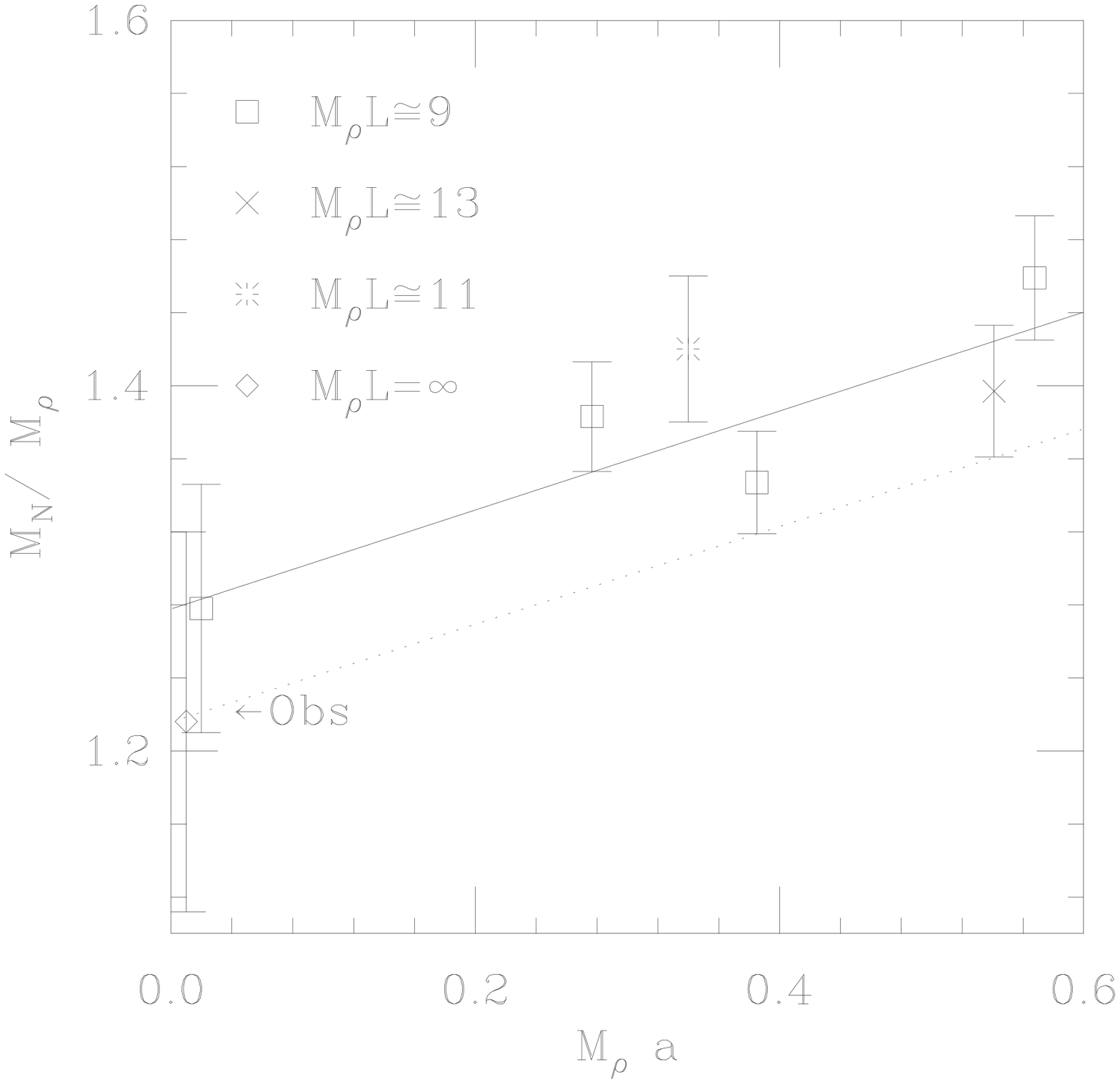}
\vspace{0.5cm}
\fcaption{
 The nucleon to $\rho$  mass ratio from Wilson fermions in the quenched
approximation~{\protect\cite{gf11,LANL}} is plotted versus lattice
spacing.  The extrapolation to the continuum limit ($a=0$) is shown as a
solid line. The lower dotted curve comes from an additional
extrapolation  in spatial size $L$ to infinite spatial volume and is
shown to agree with  the observed ratio (Obs).}
 \end{figure}

Studies have been made of the mesons and baryons which are composed of
light and strange quarks. The surprise is that the quenched
approximation  seems to reproduce~\cite{gf11} these experimental values
quite well. This may  reflect the relatively large errors that are still
present in the  lattice determinations. It may also reflect the fact
that the hadronic  dynamics has a similar energy scale in each case so
that the quenched  approximation makes similar errors --- which cancel
in mass ratios. Some possible discrepancies have been reported, however.
One is the dimensionless ratio of $m_{K^*}
(m_{K^*}-m_{\rho})/(m_{K}^2-m_{\pi}^2)$ which  comes out about 20\%
too small~\cite{lam,sinclair} in quenched evaluations. Another is the
ratio  of $b\bar{b}$ splittings $(1P-1S)/(2S-1S)$ which is found to
be~\cite{perm,nrqcd} about 10\% low.

\section{Running Coupling}

 The QCD coupling constant $\alpha$ is a key quantity in looking  for
physics beyond the minimal Standard Model. The perturbative  expansion
of QCD is accurate for energy scales beyond 10 GeV where the variation
(running) of the coupling with energy to even higher energies  is given
by the perturbative beta-function.
 $$
 {d\alpha \over dq }= -{\beta_0 \over 2 \pi} \alpha^2 -
{ \beta_1 \over 8 \pi^2}\alpha^3 - \dots
 $$
 Although lattice methods only allow a direct study of the coupling up
to scales of order 10 GeV, I shall follow convention  and quote values
of $\alpha$ at $q=m_Z$  using perturbative evolution (in the $\MSB$
scheme) to that energy.  Different regularisation schemes  lead to
different values of $\alpha$. These can be related perturbatively.  For
instance schemes `A' and `B' are related by the `matching' formula
 $$
  \alpha_A(q) = \alpha_B(q)+c_1 \alpha_B^2(q) +c_2 \alpha_B^3(q)+ \dots
 $$
 where $c_1$ and $c_2$ can be calculated perturbatively.

 Experimental determinations  of $\alpha$ inevitably need  modelling of
non-perturbative effects since  hadrons rather than quarks and gluons
are actually observed. Because  of this, it is attractive to use the
best non-perturbative method available  when determining $\alpha$:
namely lattice QCD.

In summaries of determinations of $\alpha_{\MSB}(m_Z)$, the
lattice results have relatively small errors. Several
lattice groups have produced  results. At Lattice 1993 a value of 0.108(6)
was quoted~\cite{aida} whereas the NRQCD group   quoted~\cite{nrqcd}  a
value of 0.115(2) at  Lattice 1994 and 1995. Recent
summaries~\cite{cmlat94,pw} of lattice determinations gave 0.112(7).

These values are close to the current world average value~\cite{yam} of
0.117(6).

Because the lattice method is potentially the most accurate, I discuss
in some detail the possibilities and limitations of lattice determinations
of $\alpha$.

\subsection{Lattice perturbation theory}

 The lattice formalism specifies the bare gauge coupling constant (since
the calculations fix the parameter $\beta=6/g^2$). So at $\beta=6.5$,
one has  $\alpha=g^2/(4 \pi) = 0.07346$. This is a small value of the
bare  coupling and it was expected that the perturbation series derived
from the lattice Lagrangian would be well behaved at this $\beta$-value.
However, it is now known that the perturbation series in the bare
lattice coupling converges very poorly. For instance the series  for the
logarithm of the plaquette action  $S_{\Box}$ has been calculated
analytically~\cite{pisa}  for $N_f=0$ to order $\alpha^3$ for the Wilson
gauge action.   Indeed the first eight terms of this perturbation series
have been evaluated  by a numerical technique~\cite{parma}. For SU(3)
colour,
 $$
\alpha_{\Box}  \equiv  -{3 \over 4 \pi} \log(S_{\Box})
 =  \alpha(1+3.37 \alpha + 17.69 \alpha^2 +\dots)
 $$
 \noindent Now, for $\alpha=0.07346$,  the right hand expression
evaluates to 92\% of the left hand expression obtained from the
measured lattice plaquette action. The $\alpha^3$ term contributes 7\%.
So we do not have a good convergence of perturbation theory. Decreasing
the bare $\alpha$ by a substantial factor would imply a reduction of the
lattice spacing by a huge factor which is not feasible with
present computing techniques while keeping the spatial volume a
reasonable physical size. Another way forward is needed.

The root of the poor convergence of lattice perturbation theory is
thought to come from tadpole diagrams~\cite{lm}. The way forward has
been known for a long time. A {\it physical  definition} of $\alpha$
should yield a perturbative series with  coefficients which are much
smaller than in the case of  the bare  coupling. This discussion of
improved lattice  definitions of couplings is very analogous to the
argument that the $\MSB$ scheme is preferred to the MS scheme in the
continuum. Several  improved definitions of $\alpha$ have been proposed
which are appropriate to a lattice:

(i) using the potential $V(R)$ between heavy quarks through
either $\alpha_V$ determined~\cite{lm} from $V(q)$ or
$\alpha_{F}$ determined~\cite{force} from the force $dV/dR$,

(ii) using the logarithm of the plaquette action as a
definition~\cite{lm} of $\alpha_{\Box}$,

(iii) using variation of the boundary conditions yielding $\alpha_{SF}$
from the Schr\"odinger functional~\cite{sf} and $\alpha_{TP}$ from
twisted boundary conditions~\cite{twist}.

\begin{figure}[th]

\vspace{12cm} 
 \includegraphics{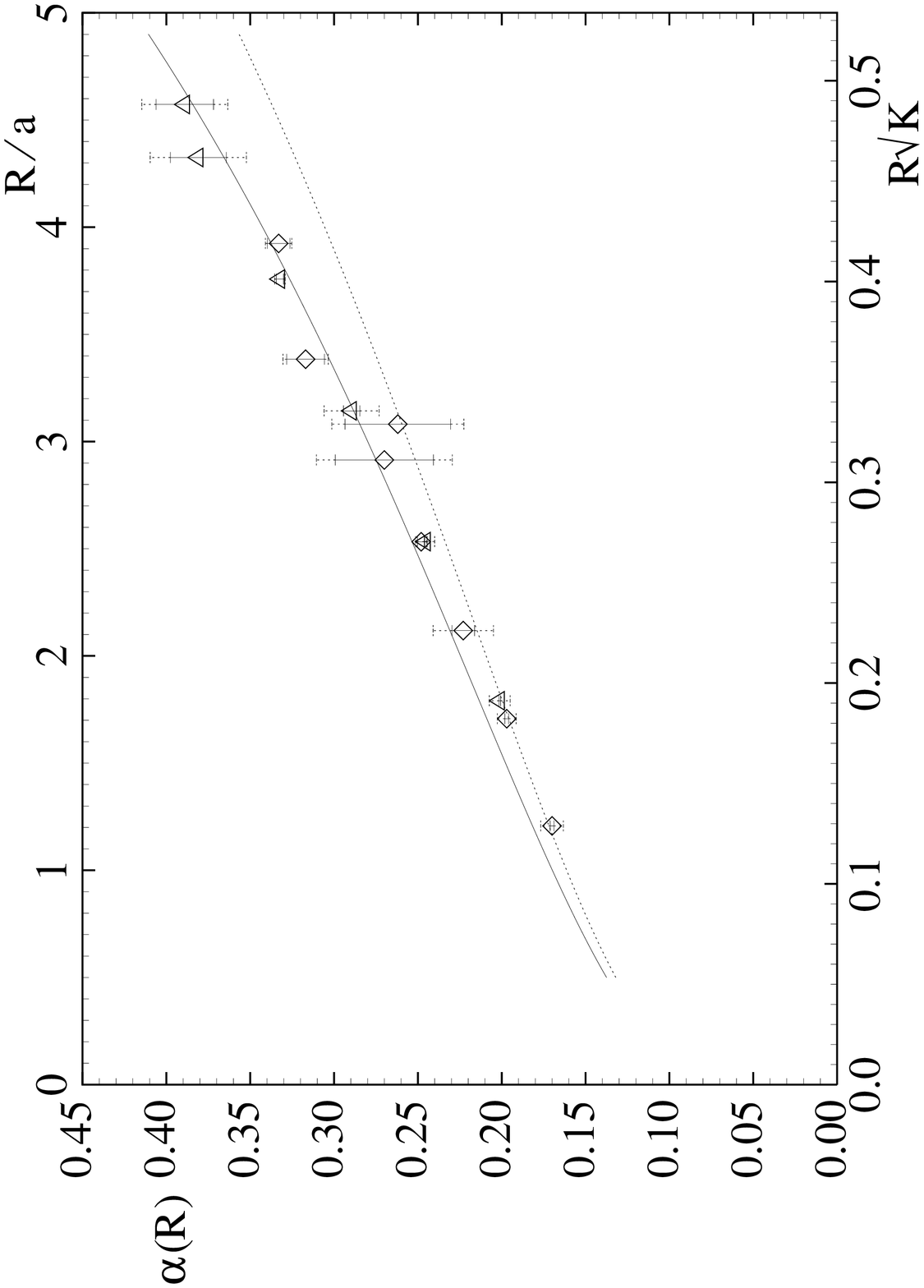}
\fcaption{
 The effective running coupling constant $\alpha_F$
obtained~{\protect\cite{ukqcdpot}} from  the force between static quarks
at separation $R$ in the quenched approximation. The scale is set by the
string tension $K$ and corresponds to energy $q \approx4$ GeV to the
left and $q \approx 1$ GeV to the right.  The curves show the two-loop
perturbative evolution for two different values of $\Lambda$. The
perturbative evolution  is not expected to be valid for $q \approx 1
$GeV.
 }
\end{figure}

As an illustration, fig~4 shows that $\alpha_F$ determined at
$\beta=6.5$ from  the potential~\cite{force,ukqcdpot} runs with energy
in a way  consistent with perturbation theory to 2 loops and has a value
which is very different from the bare $\alpha$ of 0.073.  Thus the bare
lattice coupling is seen to be anomalously small  which explains why it
is a poor expansion parameter.

 What needs to be shown is that  different improved definitions  of
$\alpha$ on a lattice agree with each other when the perturbative
matching is used. Direct tests of this are possible using quenched
SU(2) where studies using very small lattice spacings and, hence,  large
energies have been feasible.  A comparison of $\alpha_{SF}$ with
$\alpha_{TP}$ shows~\cite{twist} that, provided the scales are set
appropriately, the results agree with each other and with two  loop
perturbative evolution.  It is important to calibrate the
$\alpha_{\Box}$  determination against other methods since this  is the
method that will be used in the full QCD results discussed later. The
result~\cite{sf} is that  $\alpha_{SF}=0.1675(35)$ at
$q=1/a(\beta=2.85)$  and, since the two-loop matching is
known~\cite{lw}, one can use $\alpha_{\Box}$ to evaluate $\alpha_{SF}$
at the same scale obtaining 0.1674(5). This excellent agreement between
two different  definitions is an indication that perturbation  theory on
a lattice is now under control.

A similar comparison~\cite{cmlat94} can be made between lattice
determinations of  $\alpha_F$, $\alpha_{SF}$ and $\alpha_{\Box}$ for
quenched SU(3) and excellent agreement is obtained.

\subsection{Lattice determinations of the coupling}

 It is straightforward in principle to determine $\alpha$ on a lattice:
one uses one of the physical definitions to measure $\alpha$ at a
scale determined by a non-perturbative measurement of a quantity that
is well known experimentally (eg. the $\rho$ mass, charmonium
energy splittings or bottomonium energy splittings). It is important
to verify that the result is stable as the lattice spacing is reduced.

 At present the main limitation to this programme is that full QCD
lattice simulations have only been conducted at rather coarse
lattice spacing. The NRQCD group overcome this by using an effective
Lagrangian which is claimed to be accurate at such lattice spacings.
They extrapolate results at $N_f=0$ (ie quenched) and $N_f=2$ to
the required $N_f=3$. They set the scale using several charmonium
and bottomonium energy splittings and find consistent results~\cite{nrqcd}
for $\alpha_{\Box}$. Using one-loop perturbative matching, they
obtain $\alpha_{\MSB} =0.115(2)$.

 A more direct route to $\alpha$ uses Wilson fermions to  determine the
$\rho$ mass and charmonium energy levels. Again an  extrapolation in
$N_f$ is needed. The Kyoto-Tsukuba group also make~\cite{aoki}   an
extrapolation in the  sea-quark mass to physical sea-quark masses and
their final  result is  $\alpha_{\MSB} =0.111(5)$.

Each approach determines the coupling via $\alpha_{\Box}$ which
introduces a perturbative error of $\pm0.002c_2$ from the (as yet)
unknown two-loop  matching coefficient $c_2$. The NRQCD method uses an
effective Lagrangian  with   coefficients obtained by perturbative
evaluation, so it is hard to  estimate the systematic errors arising
from  this perturbative approximation. In the more direct approach, the
order $a$ effects can result in a scale error of 20\% on the lattice
which converts into 4\% at $m_Z$. The incomplete treatment  of the
sea-quarks in both mass and number of flavours also  introduces a
further systematic error.  Thus a conservative
summary~\cite{cmlat94,pw} of current lattice determinations is
$\alpha_{\MSB} =0.112(7)$.

It should be emphasized that, compared to other methods, the
lattice determination of $\alpha$ has no model assumptions. With
further work on (i) the two-loop matching, (ii) varying the
sea-quark mass, (iii) exploring the $N_f$ dependence and (iv) reducing
the lattice spacing $a$, it should be possible to reduce the
error on $\alpha$ to 1-2 \%.

\section{Hadronic matrix elements}

 Hadronic matrix elements which can be expressed as Green functions  can
be determined from first  principles using lattice techniques.  There
is an extensive literature on this subject~\cite{weak,soml94} and I
shall  select only some topics. Most current studies are in the quenched
approximation which introduces a source of systematic error. Even  with
this quenching error, lattice results are less model-dependent  that
most other theoretical approaches. Furthermore, it is clear how  to
remove this  quenching error by full QCD lattice studies.

One area of importance phenomenologically is   matrix elements of
electro-weak operators between hadronic states containing heavy (ie charm
or bottom)  quarks. These hadronic matrix elements are needed to relate
the quark  couplings (CKM matrix elements) to the experimental data.
They  are also needed to establish whether rare decays agree with the
Standard Model. Rare decays such as $B \to K^*\, \gamma$ involve  loops
with heavy intermediate particles and are thus prime candidates  for
tests of the Standard Model.

 There are two approaches to studying heavy quarks of mass $m_Q$ on a
lattice. One  can treat the quarks as infinitely heavy, the static
approximation, which  is very straightforward to implement since the
quarks become  effectively just  colour sources. Alternatively, one can
use propagating quarks. For  propagating quarks, one needs $m_Qa \le 1$
so that the quarks do not `fall through the gaps in the lattice'. This
latter requirement  currently implies $m_Q \le 3$ GeV, although an
improved  fermionic discretisation (such as the SW-clover formalism)
should  allow $m_Q \approx 3$ GeV.

 Since the $b$-quark lies outside each of these approaches, one needs
some theoretical input to bridge the gap.  This is
provided~\cite{neubert} by the  Heavy Quark Effective Theory (HQET)
which establishes an expansion  in $1/m_Q$ about the static limit.
These $1/m_Q$ terms can be explored  non-perturbatively on a lattice by
combining  studies of the quenched approximation with  propagating quark
results.

A different approach to heavy quarks is provided by the effective
Lagrangian of non-relativistic QCD (NRQCD). This has been used
successfully for spectroscopy~\cite{nrqcd} but has yet to be used
extensively for  matrix element calculations of the kind to be described.

 Returning now to the Heavy Quark Effective Theory, in the  heavy quark
limit, matrix elements involving heavy quarks depend only on   the
4-velocity  $v$  of the heavy quark and not on the spin or flavour.
Apart from radiative corrections arising from matching the effective
theory  to full QCD, away from the heavy quark limit there will be
corrections of order $\Lambda_{QCD}/m_Q$. The lattice  allows these
correction terms to be explored and measured. Particularly  for charm
quarks, the corrections are potentially large and a  non-perturbative
study (such as the lattice) is needed to determine  them reliably.

\subsection{ Heavy $\to$ heavy transitions}

 Semi-leptonic decays such as $\bar{B} \to D l \bar{\nu}$ are typical
applications.  In the heavy quark limit, the decays  are described by
the universal form factor $\xi(\omega)$ (Isgur-Wise form factor) which
depends only on the scalar $\omega=v.v^{\prime}$, where $v$ and
$v^{\prime}$ are the  initial and final heavy-quark 4-velocities. At the
zero-recoil point  ($\omega = 1$), the form factor is normalised by
$\xi(1)$=1 up to perturbatively-calculable radiative corrections. The
corrections of first order in $\Lambda/m_Q$ are absent at this point
which facilitates comparisons using heavy quarks of finite masses. The
HQET has no specific  prediction to make of the value of the form factor
away from this zero-recoil point. Lattice studies allow  this
to be determined --- see fig~5. Here the shape of the  form
factor as determined from lattice work~\cite{ukqcdiw} is  compared with
the distribution from  experimental data~\cite{cleoiw}.  The
distributions  are seen to have shapes which are consistent.  Thus the
ratio of the hadronic matrix element determined on a lattice to the
experimental data can be obtained  for a range of $\omega$-values  which
determines the CKM matrix element $|V_{cb}|$ relatively precisely.

\begin{figure}[th]

\centerline{\psfig{figure=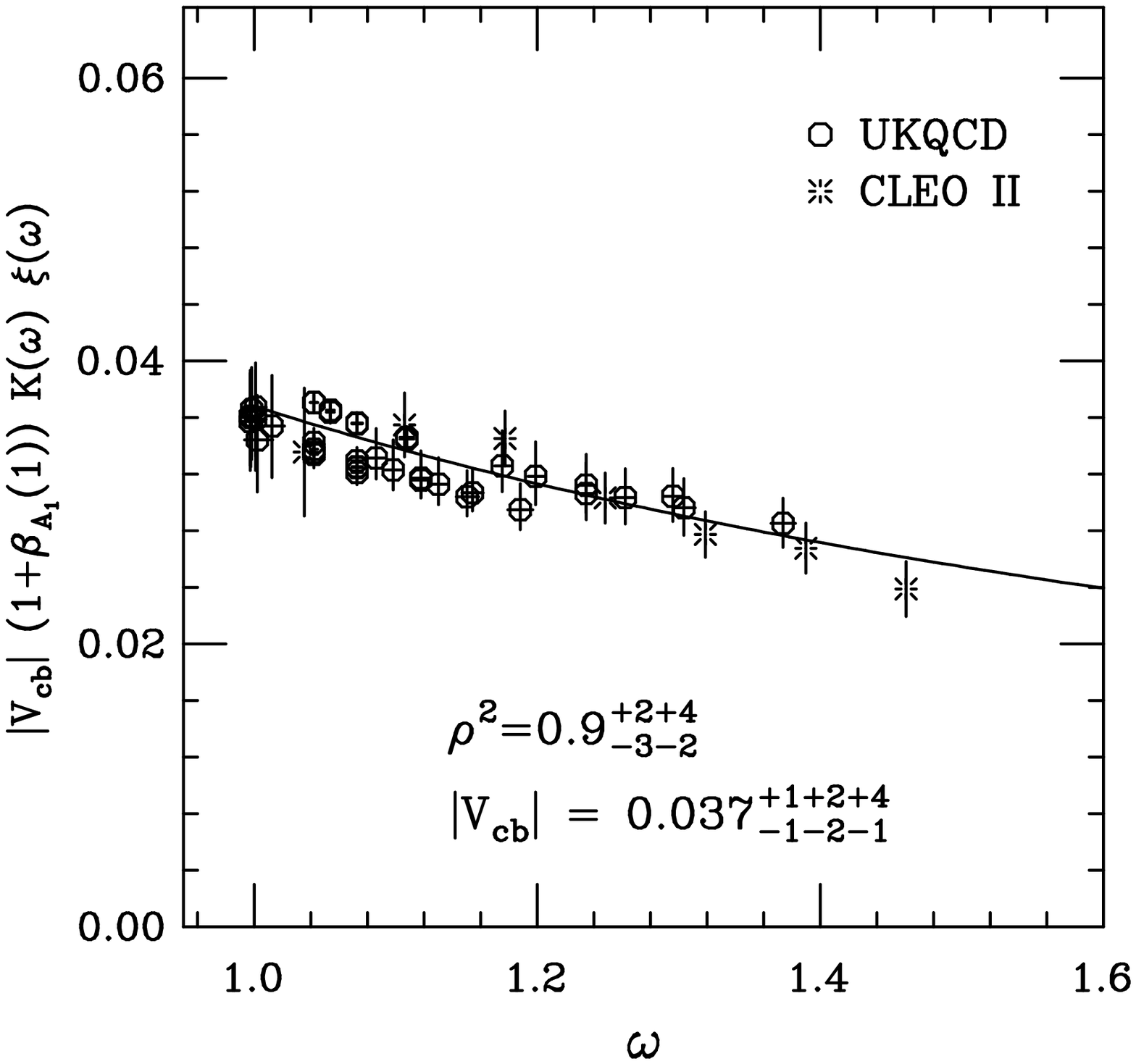,height=11cm}}
\fcaption{
  The variation with $\omega=v.v^{\prime}$ of the amplitude for the
semi-leptonic  decay $\bar{B} \to D^* l \bar{\nu}$. The lattice
results~{\protect\cite{ukqcdiw}}  and experimental
data~{\protect\cite{cleoiw}} are compared. They are seen to  have
similar shapes so that the CKM matrix element $|V_{cb}|K$ can  be
determined (where $K$ is an multiplicative radiative correction factor
of around 0.93).
 }
\end{figure}

For transitions between baryons each containing a  heavy quark, the
heavy quark limit will again be given by  a universal function. This
function will be different  from the mesonic case. A study has been made
of the baryonic Isgur-Wise function  appropriate for  semi-leptonic
transitions $\Lambda_Q \to \Lambda_{Q^{\prime}}$.  A preliminary attempt
to determine  this on a lattice has recently been made~\cite{dgr}. The
baryonic form factor  shows a faster fall-off with $\omega$ than the
mesonic case.  Applications  of this result can be made to semi-leptonic
decays such as $\Lambda_b \to \Lambda_c l \bar{\nu}$.

\subsection{ Heavy $\to$ Light transitions}

 {\it Pseudoscalar decay constants. \ } Pseudoscalar heavy-quark mesons
B and D have decay constants defined analogously to $f_{\pi}$ and  $f_K$
(where we adopt the normalisation convention that $f_{\pi} = 132$ MeV).
Since the direct leptonic decays of B and D are not measured accurately
(they  have very small branching ratio), a theoretical determination is
needed.  Again the lattice results can be analysed effectively by using
the framework of the HQET. In the heavy quark limit  of $m_Q \to
\infty$, the mass $M_P$  of the pseudoscalar meson satisfies $M_P/m_Q
\to 1$. The decay  constant for this meson  has the behaviour that $f_P
\sqrt{M_P}$ is a constant up to radiative corrections and  has  an
expansion in powers of $1/M_P$ about this heavy quark limit.

Using lattice results from static quarks and propagating quarks, this
combination can be determined for a range of values of $1/M_P$. Typical
results are shown in  fig~6. A  consistent  description of the results
is obtained using  a quadratic expression in $1/M_P$. The importance of
leading  and next-to-leading (in $\Lambda_{QCD}/M_P$) corrections to the
HQET for  this quantity were something of a surprise. Now that the
non-perturbative  lattice results are well established, the lattice data
 allow a reliable interpolation to the B meson mass. The D meson  case
is determined more directly from the propagating quark results alone.
The light quark masses used can be varied which enables results for D,
D$_s$, B and B$_s$ to be obtained.

\begin{figure}[th]
\centerline{\psfig{figure= 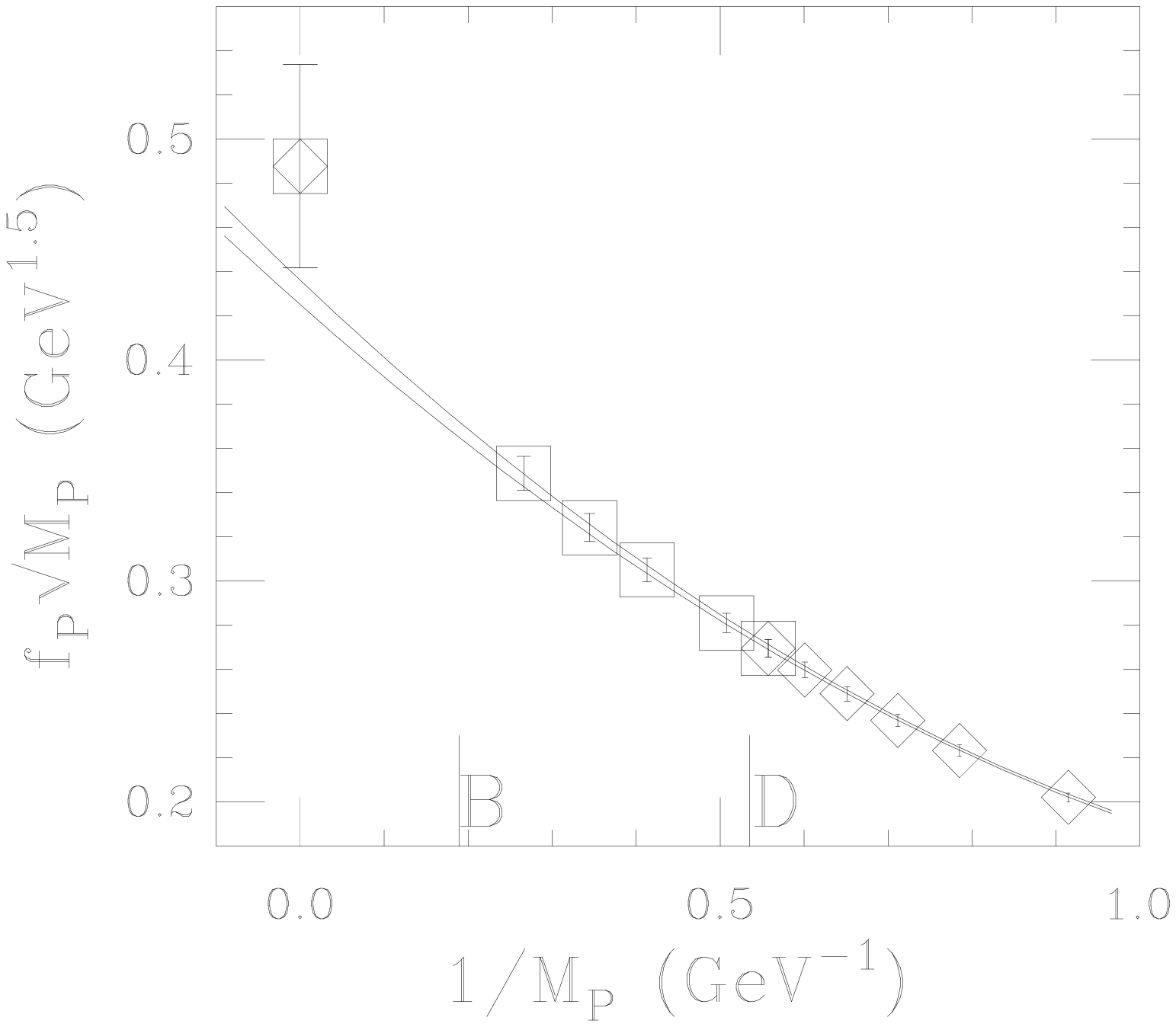,height=11cm}}
\fcaption{
  The pseudoscalar decay constant $f_P$ at different heavy quark masses
from quenched lattices~\cite{berl95}. The combination $f_P \sqrt{M_P}$
is plotted versus the inverse  pseudoscalar meson mass $1/M_P$ since
this quantity has a finite limit as $m_Q \to \infty$ in the Heavy Quark
Effective Theory. The curves show quadratic fits to the propagating quark
results. These curves extrapolate close to the static value shown.
The B and D meson mass values are marked on the scale.
 }
\end{figure} 

In any lattice determination of matrix elements there are several
sources of uncertainty. The continuum limit needs to be obtained  by
extrapolating ratios of observables to $a=0$ as was the case for mass
values.  In quenched studies, there remains an overall scale
uncertainty. The new feature is that finite renormalisation factors $Z$
are needed to  take account of the differences of the lattice action and
the  continuum action. These can be calculated perturbatively. In some
cases, however,  the first order perturbative estimate for $Z$ gives a
relatively large correction so that higher order corrections could be
significant.

However,  the lattice is a non-perturbative method and it would be
preferable  to obtain continuum matrix elements without recourse to
perturbative  approximations. Non-perturbative determination of the
relevant $Z$  factors is feasible~\cite{vlad}. One  method makes use of
Ward identities and  another uses explicit gauge fixing to define quark
and gluon  Green functions.  This is currently an area of active study
and  a more complete understanding of the systematic errors on estimates
of the $Z$ factors will be forthcoming.

{}From a conservative approach to extracting the continuum limit,
Sommer~\cite{soml94} concluded that the  quenched lattice determination
of the  heavy quark pseudoscalar decay constants is
 $$
 f_D=176(34) \ {\rm MeV} \ \ \ \ \ f_B=180(48) \ {\rm MeV}.
 $$
 The major part of the error quoted comes from the systematic errors
(continuum limit, $Z$ factors and scale). An average  of more recent
lattice results by Allton~\cite{allton} gives $ f_D \approx f_B =200 \pm
20 $ MeV.  The lattice studies are  conducted with a range of light
quark masses so that the dependence on  light quark mass can be
determined.  The ratios of $f_{D_s}/f_D$ and $f_{B_s}/f_B$ are
independent of systematic  errors on the $Z$-factors and the scale and
are found to be  between 1.1 and 1.2 in most lattice
studies~\cite{weak}.

 As yet there have been only limited studies going beyond the quenched
approximation. One recent result~\cite{berl95} obtains a value for $f_B$
some 40 MeV higher than the quenched value at a similar  lattice spacing.
Since the approach to  the continuum limit could be different for
quenched and full QCD, this full  QCD result at a relatively coarse
lattice spacing is not necessarily in  contradiction with the quenched
continuum result. A further study  of quenched QCD at several lattice
spacings is needed to obtain definitive results with all  systematic
errors under control.

  $B \bar{B}$ {\it mixing. \ } The $B$-parameter describes the hadronic
four-quark matrix element appropriate to $B \bar{B}$ mixing. This  can
be determined from lattice studies. A recent
calculation~\cite{ukqcdstat}  using the static approximation obtained
 $$
  B_B=0.99(6)(3)  \ \ \ {\rm and}\ \ \ B_{B_s}=1.01(5)(3)
 $$
 where the second error is the systematic error coming  from the scale.
There are also additional systematic errors associated  with the
matching coefficients ($Z$-factors), with extrapolation  from the
static limit of infinite quark mass to the B mass (ie corrections of
order $\Lambda_{QCD}/M_B$) and with quenching. The lattice results  for
$B_B$ are close to  the naive value of unity for this matrix element.

The combination $f_B^2 B_B$ enters $B \bar{B}$ mixing. Since the
errors on $f_B$ coming from the $Z$ factors and scale are relatively
large, it is useful to take the ratio between strange and non-strange
B mesons to have a better determined quantity. Thus
 $$
{f_{B_s}^2 B_{B_s} M_{B_s} \over f_{B_d}^2 B_{B_d} M_{B_d}} = 1.37(9)(5)
 $$
 from a recent quenched lattice evaluation~\cite{ukqcdstat} in the
static limit. This ratio allows the observed $B \bar{B}$ mixing
coefficients $\Delta m_s/\Delta m_d$ to be  related to the CKM matrix
element ratio $(V_{ts}/V_{td})^2$. The  present experimental
situation~\cite{slwu} is that only a lower bound exists on $\Delta m_s$
(namely $\Delta m_s > 9 \tau_{B_s}$) and this gives constraints on the
CKM matrix elements which are consistent with other  information.

Conversely, one can use existing knowledge  of CKM matrix  elements and
the lattice estimates of  the hadronic matrix elements to evaluate a
prediction for $x_s = \Delta m_s /\tau_{B_s}$. Assuming $B_K=0.8$  and
$f_B=180$ MeV, a rather large value of $x_s \approx 16$ is
indicated~\cite{ukqcdstat}  and such a large value would render mixing
effects for $B_s$ very hard to detect  experimentally.

 {\it Rare B decays.\ }  Decays such as $B \to K^{*} \gamma$ are
sensitive tests of the Standard Model since they are forbidden  at
tree-level (as a flavour changing neutral current) and only proceed via
loops  with heavy intermediate states (W and t). One of the relevant
contributions comes from the penguin diagram. The short-distance  part
can be calculated perturbatively but  the hadronic matrix element
between  B and K$^*$ and the appropriate four-quark operator is needed.
This  can be calculated in principle by lattice methods.  At present
this is a  rather exploratory study and results are not yet fully
established. Nevertheless,  as an example of the potential of the
lattice technique, I discuss the  possibilities.

  The appropriate hadronic matrix element can be calculated using
propagating heavy quarks on the lattice. This does not allow to reach
the B-meson mass without an extrapolation. For an improved fermionic
action such as the SW-clover action, the range of mass values (or
momentum values) of the heavy quark  that can be reliably used is larger.

 In order to establish a theoretical framework for the extrapolation in
heavy quark mass, one can use the HQET which  implies that physics  is
similar at a fixed value of
 $$
\omega=v.v^{\prime}=1+{(M-m)^2-q^2 \over 2Mm}
 $$
 where $M$ is the heavy Meson mass and $m$ the  K${^*}$ mass. For the
decay $B \to K^{*} \gamma$, the photon is on-shell with $q^2=0$ and thus
$\omega =3.04$. This large value of $\omega$  implies, at a lower value
of $m_Q$, a large value of $q^2$.  Present  lattice studies are not able
to explore this region of $\omega$ directly. Models  for the $q^2$
dependence of the matrix element can be constructed and  checked against
measurements in the available $q^2$ interval. A  range of plausible
extrapolations can be found~\cite{rare}, but they lead to a rather wide
band of predictions for the decay rate.   One way to focus on the
hadronic matrix element is to study the  branching ratio to the
inclusive radiative decay to strange mesons $(B \to K^* \gamma) /(b \to
s \gamma)$.  This branching ratio is $19\pm6\pm4$ \%
experimentally~\cite{skwarnicki}  and values straddling this are
obtained~\cite{rare} from different extrapolation  assumptions from
lattice data.

Further progress needs a reduced lattice spacing to allow larger  quark
masses and/or large momenta to be explored directly. The HQET relates
the  hadronic matrix elements for the rare decay  $B \to K^{*} \gamma$
to those for the semi-leptonic decays $B \to L l\bar{\nu}$ and $D \to L
l\bar{\nu}$ where $L$ stands for a light meson: $\pi$, $\rho$, K or
K$^*$. Some  theoretical input can also come from a combined study of
these related processes. Here I stress one such application.

\begin{figure}[h]
\centerline{\psfig{figure= 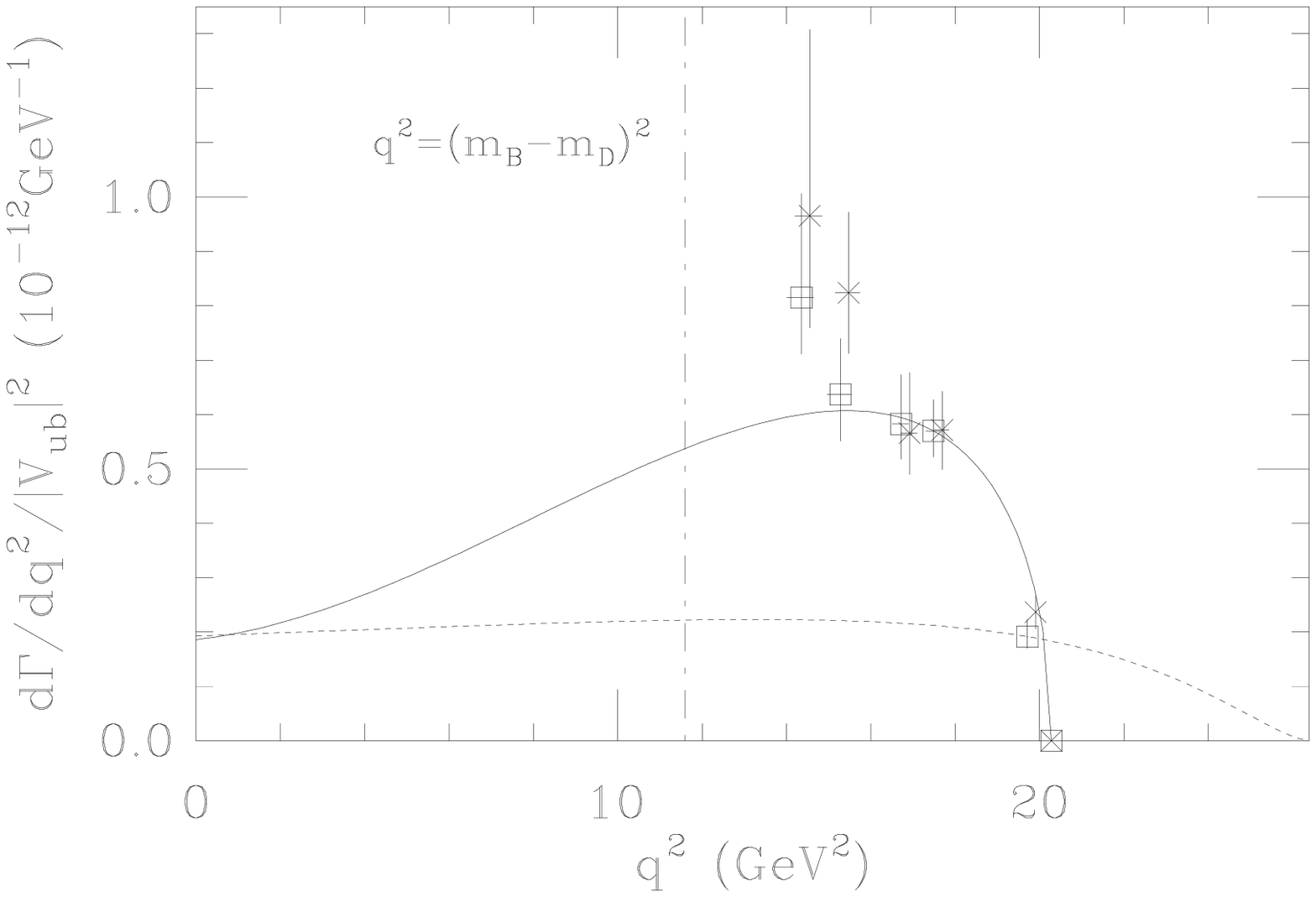,height=11cm}}
\fcaption{
  The distribution versus $q^2$ of the rate for the decay  $\bar{B^0}
\to \rho^+\, l^-\,  \bar{\nu}$. The lattice
results~{\protect\cite{brho}}  for  the hadronic matrix element are
shown together with a continuous curve  which is a plausible estimate
of the expected shape of the distribution.  The vertical dash-dotted
line represents the upper limit in $q^2$  for charm production.  The
dotted curve illustrates a plausible  shape for the decay
$\bar{B^0} \to \pi^+\, l^-\,  \bar{\nu}$.
 }
\end{figure} 

{\it $B \to \rho +l +\bar{\nu}$}.\ \ \
 Exclusive semi-leptonic decays of B mesons to non-strange light hadrons
give a  route to determine the $V_{ub}$ CKM mixing strengths. As well as
accurate experimental data on branching ratios, the hadronic  matrix
element needs to be evaluated from QCD accurately to allow  this
determination. This has recently been addressed from  lattice
measurements.

Currently one can study on the lattice the relevant matrix elements in
the region $m_Q \approx m_c$ and $q^2   \raisebox{-.75ex}{ {\small
\shortstack{$<$ \\ $\sim$}} }   m_c^2$.    Using again  HQET to control
the extrapolation in $m_Q$ to $m_b$, implies that the  region explored
on the lattice corresponds~\cite{brho} for B decay to the region  of
phase space where  the momentum transfer $q$ from the heavy quark (Q) to
the light meson ($\rho$)  is near its maximum. This is  illustrated in
fig~7. This is a phase space region which is expected  to contribute to
a substantial fraction of the decay events. Moreover, this region is
above the endpoint for charm production  in the decay and  thus there
will be no contamination  from the larger branching fraction to charm
final states. As experimental data on this exclusive  decay becomes more
precise, the comparison of experimental and lattice distributions  in
this interval at  large $q^2$ will provide an accurate  method to
determine $|V_{ub}|$.

\section{Conclusions}

The lattice technique is now a well established and quantitative method
to obtain {\it continuum} results in quenched QCD.  I have discussed
applications to glueballs, mesons and determination of $\alpha_S$.
Other topics, only partly discussed here,  include mesons
and baryons involving one heavy quark,  hybrid mesons and the $c
\bar{c}$ and $b \bar{b}$ systems. As well as masses, decay matrix
elements, form factors and  moments of structure functions can be
evaluated. Another important area of study which relies heavily on
lattice input is hadronic physics at  non-zero temperature. There are
limitations to the current lattice  capabilities: only quantities which
can be expressed as vacuum expectation values of fields can be explored
easily. Consequently, for example, (i) non-zero chemical potential is
hard to explore, and (ii) scattering can only be studied indirectly and
in a limited regime.

The bigger limitation is the quenched approximation itself. No
comprehensive studies have been made of the corrections caused by
quenching.  Most of our experience comes from results for two flavours
of sea quarks that are too heavy and with a rather coarse lattice
spacing.  This situation can be improved either by increased computing
resources  or by improved  algorithms. The lattice community is working
hard on both fronts.  At present 10 -- 50 Gflops-year is the typical
resource available to Grand Challenge lattice projects (this  is
$\approx 10^{18}$ floating point operations). There are well-advanced
plans for computing resources providing 100 -- 500 Gflops dedicated to
lattice  work.  The increase in computational power will allow fuller
studies  of the systematic errors coming from quenching.

Meanwhile, a way to  explore beyond the quenched approximation is to
evaluate explicit quark loops  in the quenched approximation.  Several
groups have explored this route  and have been able to address topics
such as the spin content of the  nucleon, the $\eta$--$ \eta'$ system
and hadronic decay widths. Estimates  of hadronic scattering lengths
have also been given.

A fuller understanding of the determination of hadronic matrix elements
is in progress. This is important for the particle physics community  as
a whole, since accurate values of such matrix elements are an essential
component of precision tests of the Standard Model and searches for
physics beyond the Standard Model.

\end{document}